\documentclass[twocolumn,showpacs,preprintnumbers]{revtex4}

\usepackage{amssymb}
\usepackage{graphicx}
\usepackage{dcolumn}
\usepackage{bm}
\usepackage{latexsym}
\usepackage{float}

\begin{document}

\def\bbox#1{\hbox{\boldmath${#1}$}}
\def\gtsim{$\raisebox{0.6ex}{$>$}\!\!\!\!\!\raisebox{-0.6ex}{$\sim$}\,\,$}
\def\ltsim{$\raisebox{0.6ex}{$<$}\!\!\!\!\!\raisebox{-0.6ex}{$\sim$}\,\,$}
\def\pt{\bbox{p}_t}
\def\xx{\hbox{\boldmath{$ x $}}}
\def\pp{\hbox{\boldmath{$ p $}}}
\def\pt{\pp_{{}_T}}
\def\yb{{\bar y }}
\def\mt{m_{{}_T}}
\def\zb{{\bar z }}
\def\tb{{\bar t }}
\def\rhoe{ {\rho_{\rm eff}} }
\def\qq{\hbox{\boldmath{$ q $}}}

\title{ Productions of $Z^0$ and $W^+/W^-$ in Relativistic Heavy-Ion Collisions at the LHC }

\author{Peng Ru$^{1,2}$\footnote{Electronic address:
pengru@mail.dlut.edu.cn}}

\author{Ben-Wei Zhang$^{2}$\footnote{Electronic address:
bwzhang@mail.ccnu.edu.cn}}

\author{Luan Cheng$^{1}$}

\author{Enke Wang$^{2}$}

\author{Wei-Ning Zhang$^{1}$}

\affiliation{$^1$School of Physics $\&$ Optoelectronic Technology,Dalian
University of Technology,Dalian,116024 China}

\affiliation{$^2$Key Laboratory of Quark \& Lepton Physics (MOE) and Institute of Particle Physics,
 Central China Normal University, Wuhan 430079, China}

\received{\today}

\begin{abstract}
The productions of massive gauge bosons, $Z^0$ and $W^+/W^-$, in heavy-ion reactions at the LHC, provide an excellent tool to study the cold nuclear matter effects in high-energy nuclear collisions. In this paper we investigate $Z^0$ and $W^+/W^-$ productions in p+Pb and Pb+Pb at the LHC, at NLO and NNLO with DYNNLO incorporating the nuclear PDFs (nPDFs) parametrization sets EPS09 and DSSZ, within the framework of perturbative QCD. The numerical simulations of the transverse momentum spectra, rapidity dependence, and related nuclear modification factors for $Z^0$ and $W$ particles, as well as the charge asymmetry for W boson, are provided and tested against the latest experimental data. It is found that the theoretical results with EPS09 and DSSZ nPDFs can give good descriptions of the recent data on $Z^0$ and $W^{\pm}$ particles in p+Pb and Pb+Pb within the experimental error bars, though some differences between results with EPS09 and DSSZ can be observed, especially in the rapidity dependence of the $Z^0$ yield. Theoretical predictions for future measurements on $Z^0$ and $W$ in p+Pb and Pb+Pb collisions at the LHC are also provided.

\end{abstract}

\maketitle
\section{Introduction}
\label{introduction}

With the running of the Large Hadron Collider~(LHC), the measurement of heavy gauge bosons $Z^0$ and $W^+/W^-$ in relativistic heavy-ion
collisions has become available for the first time and attracted a lot of attentions~\cite{Aad:2010aa,Chatrchyan:2011ua,Chatrchyan:2012nt,Aad:2012ew,atlasz14,CMS:2014kla,CMS:2014sca,Aaij:2014pvu,Aad:2014bha,Chatrchyan:2014csa}.
Through the Drell-Yan~(DY) mechanism~\cite{DY}, the gauge boson production with final state lepton pair provides an interesting insight on perturbative quantum chromodynamics~(pQCD) for
both hadronic and nuclear collisions.
It's also noteworthy that in the heavy-ion collisions~(HIC), the electro-weak boson production is hardly
affected by the evolution of the extremely hot and dense QCD matter,
since the boson (decayed lepton pair) mean-free-path in the QCD matter is much longer than the size of
the QCD medium~\cite{Kartvelishvili:1995fr,ConesadelValle:2007sw,ConesadelValle:2009vp}.
Therefore, the production of massive gauge boson is insensitive to the possible final-state interactions and provides a good probe of the initial-state cold nuclear matter (CNM) effects in high-energy nuclear collisions~\cite{Kartvelishvili:1995fr,ConesadelValle:2007sw,ConesadelValle:2009vp,Vogt:2000hp,Paukkunen:2010qg,Guzey:2012jp,Kang:2012am,Albacete:2013ei}.

In this paper we focus on the CNM effects on the massive gauge boson production and
investigate the yields of $Z^0$ and $W^+/W^-$ bosons in nuclear collisions at the next-to-leading order~(NLO)
and the next-to-next-to-leading order~(NNLO) within the framework of pQCD by using the program DYNNLO~\cite{DYNNLO}. Several important
CNM effects such as the isospin effect, nuclear shadowing effect, anti-shadowing effect, EMC effect as well as Fermi motion effect will be included
in the numerical simulations of heavy gauge boson production by phenomenologically utilizing the nuclear parton distribution functions (nPDFs) parametrization
sets EPS09~\cite{Eskola:2009uj} and DSSZ~\cite{deFlorian:2011fp}.  The study presented in this paper is a first attempt to make a complete theoretical
investigation on productions of three massive gauge boson ($Z^0$ and $W^+/W^-$) in both p+Pb and Pb+Pb collisions at the NNLO. The transverse momentum and rapidity distribution of $Z^0$ and $W^+/W^-$ yields, the $W$ boson charge asymmetry in p+Pb and Pb+Pb collisions will be computed numerically and compared with the data at the LHC. Furthermore, the difference between results with two different parametrization sets of EPS09 and DSSZ will be shown, and their physics implications will be discussed. It is noted that $Z^0$ nuclear modification factor as a function of transverse momentum is an optimal tool to study the alterations of gluon distribution in a nucleus, whereas  $Z^0$ yield as functions of the $Z^0$ boson rapidity
provides important constraints on the nuclear modifications of quark distribution functions. Due to isospin effect, $W^{\pm}$ charge asymmetry will be modified significantly in p+Pb and Pb+Pb collisions and it is found that at the same colliding energy the charge asymmetry in
p+Pb collision may approach that in p+p collision at positive charge lepton pseudorapidity, but approach that in Pb+Pb at negative charge lepton pseudorapidity.

Our work is organized as follows: In Section~\ref{section:gaugeboson@pp}, we discuss the gauge boson productions
in hadronic collisions through the Drell-Yan mechanism and introduce the program DYNNLO.
In Section~\ref{section:Z@HIC}, we study the $Z^0$ boson production
in heavy-ion collisions at LHC for both Pb+Pb at $\sqrt{s_{NN}}=2.76$~TeV and p+Pb at $\sqrt{s_{NN}}=5.02$~TeV.
We calculate the $Z^0$ boson transverse momentum distribution, rapidity distribution,
and the corresponding nuclear modification factor.
The parton-flavor dependent nuclear modification effects on the gauge boson production are discussed.
In Section~\ref{section:W@HIC}, we focus on the $W^+/W^-$ boson production in HIC.
The isospin effect on the $W$ transverse momentum distribution and the charge asymmetry are mainly discussed.
In Section~\ref{section:summary&conclusion}, we give the summary and conclusions.

\section{$Z^0$ and $W$ productions in hadronic collisions}
\label{section:gaugeboson@pp}
To consider the vector boson production in heavy-ion collisions, as a first step we should know how to compute its production in elementary hadron-hadron reactions.
Generally we consider the following inclusive hard-scattering reaction
\begin{eqnarray}
h_1+h_2\to V + X
\end{eqnarray}
where $ V=Z^0/\gamma^*$, $W^+$, or $W^-$ with
 high invariant mass. Taking $Z^0/\gamma^*$ production as an example, in the parton
model, the cross section $\sigma_{DY}$ for the Drell-Yan process
with $Z^0$ decaying to a lepton pair can be calculated
by weighting the subprocess cross section
$\hat{\sigma}$ for $q\bar{q}\to Z^0\to ll$ (at
the lowest order) with the parton
distribution functions (PDFs) of quark $q(x,\mu)$ and that of anti-quark $\bar{q}(x,\mu)$,
and then summing contributions of all quark-antiquark combinations
in $h_1$ and $h_2$ as~\cite{Fie95, QCDcp}
\begin{eqnarray}
\label{DY}
\sigma_{DY}&=&\sum_{f}\int dx_1dx_2[q_1(x_1,\mu)\bar{q}_2(x_2,\mu) \nonumber \\
&+&\bar{q}_1(x_1,\mu)q_2(x_2,\mu)]\times \hat{\sigma}_{q\bar{q}\to V(q)\to ll} \, .  \,
\end{eqnarray}
Here $V\to ll$ stands for the process of $Z^0/\gamma^*\to l^+l^-$.

In calculating the production of the massive gauge bosons,
higher order corrections to Eq.~(\ref{DY}) play an important role.
At the next-to-leading order which gives perturbative expansions at $\mathcal {O}(\alpha_s)$, there are two kinds
of new contributions: (1) the real corrections, and (2) one-loop virtual corrections to
the  leading order (LO) subprocess;
at the next-to-next-to-leading order which gives calculations at $\mathcal {O}(\alpha_s^2)$, three
kinds of contributions should be included: double real corrections, real-virtual corrections,
and two-loop virtual corrections to the LO subprocess~\cite{DYNNLO,Gie92,Fri96,Cat97,fewz}.
In the last two decades, some numerical approaches of computing massive gauge boson productions in hadronic collisions at the NLO and NNLO
have been developed~\cite{DYNNLO,mcfm,fewz}.
In this work, the Drell-Yan process cross sections of $Z^0$ and $W^{+}/W^-$ are calculated by using the NNLO
computation program DYNNLO~\cite{DYNNLO}, which includes all the higher order corrections at  $\mathcal {O}(\alpha_s)$ and  $\mathcal {O}(\alpha_s^2)$ mentioned above and also takes into account finite-width effects, $\gamma-Z^0$ interference,
the leptonic decay of the vector bosons as well as the corresponding spin correlations.

Besides considering the total yields of massive gauge bosons and their rapidity dependence, we also investigated their transverse momentum spectra.
Although the lowest order of the DY process in Eq.~(\ref{DY}) gives important contribution to the total yield, it does not give non-vanishing transverse momentum $p_T^V$. Therefore, when $p_T^V\neq 0$, the (N)NLO contributions of total yield are
given by the (N)LO contributions to the final state $V$+jet(s). So the leading-order contribution to the total cross section (or rapidity distribution)
of $Z^0$ is given by $q\bar{q}\rightarrow Z^0$, whereas for $Z^0$ transverse momentum distribution, the processes of $q\bar{q}\rightarrow V+g$ and $qg\rightarrow V+q$ provide the dominant contribution. It is also noted that in computing transverse momentum spectra
at sufficiently small $p_T^V$, the resummation technique will be needed to get robust predictions~\cite{Dav84,Col85,xzhang02,Landry:2002ix,Bozzi:2010xn}. In this work we will investigate rapidity dependent cross section of the massive gauge boson and its transverse momentum
distribution at large $p_T^V$ region with DYNNLO, which has been shown to give  descriptions on the total cross section and rapidity distribution of the vector boson as well as its large transverse momentum spectrum at Tevatron~\cite{DYNNLO}.
\begin{figure}[t]
\begin{center}
\includegraphics[scale=0.70]{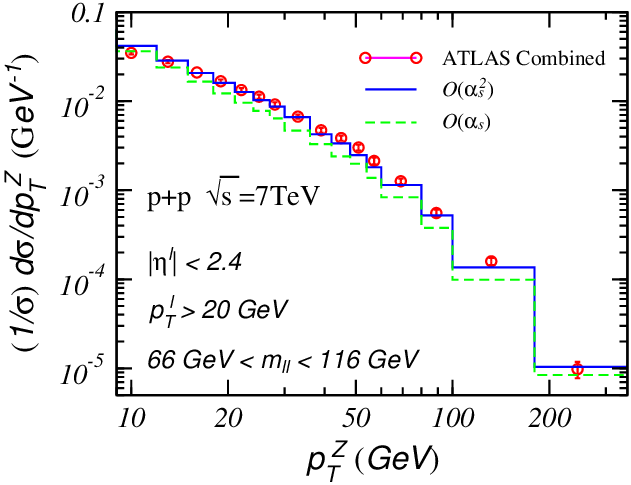}
\end{center}
\caption{(Color online) The normalized $Z^0$ boson differential cross section as
a function of $p_T^Z$ in p+p collisions at $\sqrt{s}=7$~TeV.
The ATLAS combined data are taken from Ref.~\cite{Aad:2011gj}. }
\label{fig:Z-pt@pp}
\end{figure}
\begin{figure}[t]
\label{rnlrtd}
\begin{center}
\includegraphics[scale=0.70]{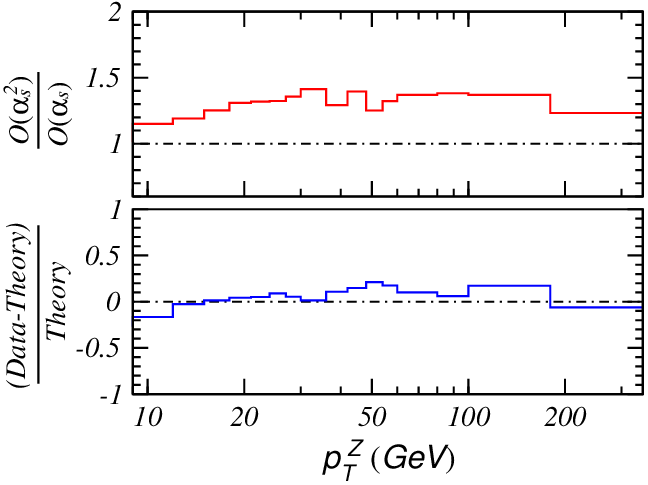}
\end{center}
\caption{(Color online) Top panel: the ratio of the normalized $Z^0$ cross section at
$\mathcal {O}(\alpha_s^2)$ to that at $\mathcal {O}(\alpha_s)$. Bottom panel:
the deviation between theory for the $Z^0$ production at $\mathcal {O}(\alpha_s^2)$ and ATLAS data~\cite{Aad:2011gj}.}
\label{fig:K-factor@pp}
\end{figure}

In this paper to compute numerically heavy gauge boson productions at (N)NLO with DYNNLO~(version 1.4),
Martin-Stirling-Thorne-Watt (MSTW)~2008~\cite{Martin:2009iq} sets of parton
distribution functions~(PDFs) are chosen, and the NNLO PDFs are used in calculations at the $\mathcal {O}(\alpha_s^2)$ and the NLO PDFs in those at the
$\mathcal {O}(\alpha_s)$. The renormalization scale ($\mu_R$) and the factorization
scale ($\mu_F$) are both fixed at the value $\mu_R=\mu_F=m_V$, with $m_V$ the mass of the vector boson.
One intensive study about the impact of the renormalization and factorization scales on the
gauge boson rapidity distribution can be found in Ref.~\cite{Anastasiou:2003ds}, and discussions
about these scales on the transverse momentum spectrum calculation
can be seen in Ref.~\cite{Aad:2011gj}. Generally, the calculations with higher order pQCD corrections have less dependence
on the scales.

The productions of massive gauge bosons in the proton-proton collisions
at $\sqrt{s}=7$~TeV have been studied by measuring their transverse momentum and
rapidity distributions. We confront the theoretical calculations with these measurements
to test the validity of perturbative QCD calculations of the heavy gauge boson productions.

In Fig.~\ref{fig:Z-pt@pp} we compare the theoretical calculations of the normalized $Z^0$ differential cross section
 $({1}/{\sigma}){d\sigma}/{dp_T^Z}$ with the ATLAS
measurement which combines the channel of $Z^0/\gamma^*\to ee$ with the channel of $Z^0/\gamma^*\to \mu\mu$.
The final state fiducial phase space is defined by the lepton pseudorapidity and transverse
momentum, and by the invariant mass of the dilepton:
$|\eta^l|<2.4$, $p_T^l>20$~GeV and $66$~GeV $< m_{ll} < 116$~GeV~\cite{Aad:2011gj}.
We see the $\mathcal {O}(\alpha_s^2)$ results agree very well with the data in the region of $p_T^Z>10$~GeV.
To see the relative contributions of higher order corrections, and the deviation between theory and experiment
we plot in Fig.~\ref{fig:K-factor@pp}
the ratios of the $\mathcal {O}(\alpha_s^2)$ results to the $\mathcal {O}(\alpha_s)$ results, and the ATLAS data
to the $\mathcal {O}(\alpha_s^2)$ results. It is observed that
the $\mathcal {O}(\alpha_s^2)$ calculations give about $10\%\sim40\%$  more contributions
than those of $\mathcal {O}(\alpha_s)$ to the differential cross sections as a function of $p_T^Z$,
and the differences between theory and data are rather small.

\begin{figure}[t]
\begin{center}
\includegraphics[scale=0.7]{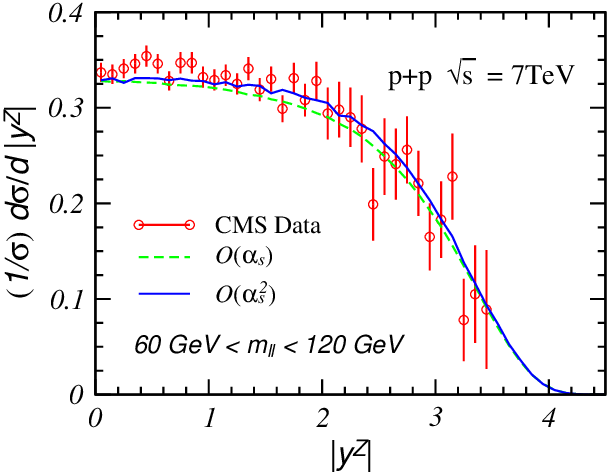}
\end{center}
\caption{(Color online)  The normalized $Z^0$ boson differential cross section as
a function of rapidity in p+p at $\sqrt{s}=7$~TeV.
The CMS data are taken from Ref.~\cite{Chatrchyan:2011wt}.}
\label{fig:Z-y@pp}
\end{figure}

\begin{figure}[h]
\label{rnlrtd}
\begin{center}
\includegraphics[scale=0.70]{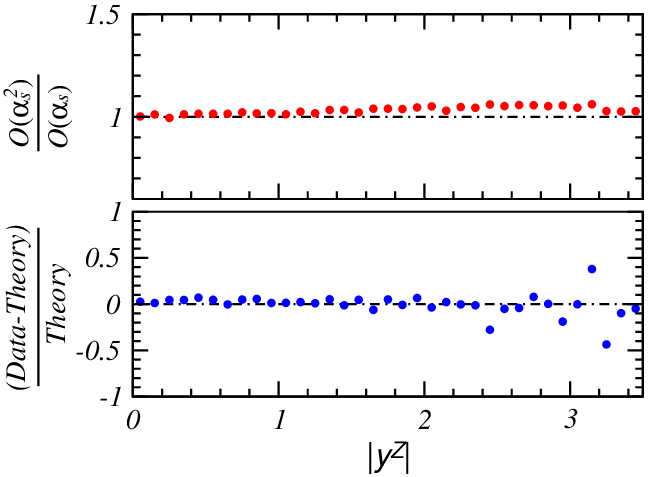}
\end{center}
\caption{(Color online) Top panel: the ratio of the normalized $Z^0$ cross section at
$\mathcal {O}(\alpha_s^2)$ to that at $\mathcal {O}(\alpha_s)$. Bottom panel:
the deviation between theory for the $Z^0$ production at $\mathcal {O}(\alpha_s^2)$ and CMS data~\cite{Chatrchyan:2011wt}.}
\label{fig:K-factory@pp}
\end{figure}

\begin{table}[tbh]
\tabcolsep 0pt \caption{The total cross sections for massive gauge bosons
in p+p collisions at $\sqrt{s}=7$~TeV.
The calculations are done for the ATLAS fiducial volume; see text for definition.
 } \vspace*{-12pt}
\begin{center}
\def\temptablewidth{0.4\textwidth}
{\rule{\temptablewidth}{1pt}}
\begin{tabular*}{\temptablewidth}{@{\extracolsep{\fill}}ccc}
{\rm Vector} & cross section(nb)  & cross section(nb)  \\{\rm Boson
} & at $\mathcal {O}(\alpha_s)$ & at $\mathcal {O}(\alpha_s^2)$ \\
\hline $Z^0$ & $0.45\pm0.0002$ &
$0.458\pm0.0008$ \\  $W^+$ & $3.000\pm0.0016$ & $3.062\pm0.0092$  \\
$W^-$ & $2.025\pm0.001$ &
 $2.045\pm0.0048$  \\ \hline
\end{tabular*}
{\rule{\temptablewidth}{1pt}}
\end{center}
\label{table:yield@pp}
\end{table}

In TABLE.~\ref{table:yield@pp} we list the theoretical simulation of the total cross sections of $Z^0$ and $W^+/W^-$ in the fiducial phase space
at the NLO and NNLO for p+p reactions at the LHC.
It is seen that the total cross sections of the gauge bosons at the NNLO will be enhanced slightly as compared to those at the NLO.
The errors of theoretical calculations shown in the table, are provided by the DYNNLO program according to an estimate of the numerical
errors in the Monte Carlo integration~\cite{DYNNLO}.

The normalized differential cross sections as a function of $Z^0$ boson rapidity
in the invariant mass interval $60$~GeV$<m_{ll}<120$~GeV are measured by CMS collaboration~\cite{Chatrchyan:2011wt}.
We show our numerical results with the CMS data in Fig.~\ref{fig:Z-y@pp}, and we find that both the
$\mathcal {O}(\alpha_s^2)$ and $\mathcal {O}(\alpha_s)$ calculations have
good agreements with the data.
We also plot in Fig.~\ref{fig:K-factory@pp}
the ratios of the $\mathcal {O}(\alpha_s^2)$ results to the $\mathcal {O}(\alpha_s)$ results, and the CMS data
to the $\mathcal {O}(\alpha_s^2)$ results.
The results at $\mathcal {O}(\alpha_s^2)$ are slightly
larger than those at $\mathcal {O}(\alpha_s)$, just like the case of total
cross sections calculations.

\begin{figure}[t]
\includegraphics[scale=0.7]{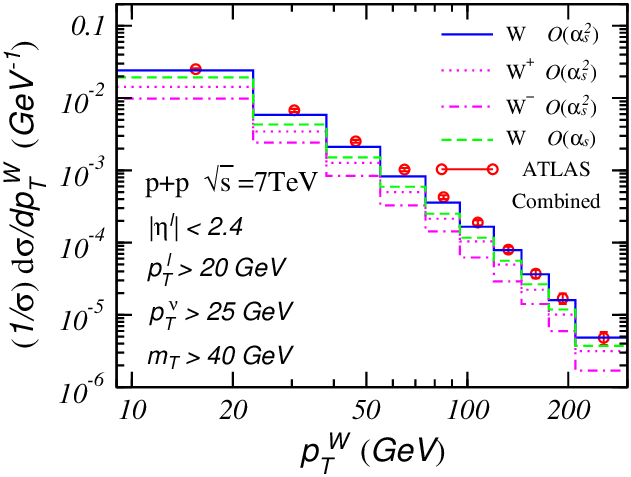}
\caption{(Color online) The normalized $W$ boson differential cross section as
a function of $p_T^W$ in p+p collisions at $\sqrt{s}=7$~TeV.
The ATLAS combined data are taken from Ref.~\cite{Aad:2011fp}. }
\label{fig:W-pt@pp}
\end{figure}
The transverse momentum distribution of $W$ bosons is also studied, and the
theoretical calculations as well as the experimental data by ATLAS are illustrated
in Fig.~\ref{fig:W-pt@pp}.
Both the theory and data are made in the fiducial phase
space, which is defined by the pseudorapidity and transverse momentum of charged
lepton, the transverse momentum of neutrino, and the transverse mass as
$|\eta^l|<2.4$, $p_T^l>20$~GeV, $p_T^{\nu}>25$~GeV,
$m_T=\sqrt{2p_T^lp_T^{\nu}(1-cos(\phi^l-\phi^{\nu}))}>40$~GeV~\cite{Aad:2011fp}.

Figures~\ref{fig:Z-pt@pp}-\ref{fig:W-pt@pp} demonstrate that the pQCD calculations of
heavy gauge boson productions at the (N)NLO with DYNNLO can give satisfactory descriptions of the
experimental data, and thus provide a very good theoretical tool to study the productions of
the gauge bosons in relativistic heavy-ion collisions and related cold nuclear matter effects.

\section{$Z^0$ boson production in heavy-ion collisions}
\label{section:Z@HIC}
To extend the investigation of the gauge boson productions in p+p collisions to that in p+A and A+A reactions,
cold nuclear matter (CNM) effects should be taken into account, even though final-state hot/dense QGP effects
(which may exists in Pb+Pb collisions at the LHC) will not affect the production of the gauge boson production due to
its much larger mean-free-path relative to the size of the QGP~\cite{Vogt:2000hp,xzhang02,Neufeld:2010fj,Paukkunen:2010qg,Dai:2013xca,Dai:2012am,Albacete:2013ei,Kang:2012am}.
To include several CNM effects (such as shadowing, anti-shadowing, EMC effect, Fermi motion {\it etc.}) on heavy gauge boson
production in computing the cross sections in heavy-ion collisions, we replace the parton distribution functions (PDFs) of the free proton with the nuclear parton distribution functions (nPDFs)~\cite{Eskola:2009uj,deFlorian:2011fp}. We note that the effect of parton energy loss in cold nuclear matter~\cite{Schafer:2007xh, Neufeld:2010dz, Xing:2011fb} has not been included in this approach and we may defer it in a future study.

\begin{figure}[t]
\includegraphics[scale=0.7]{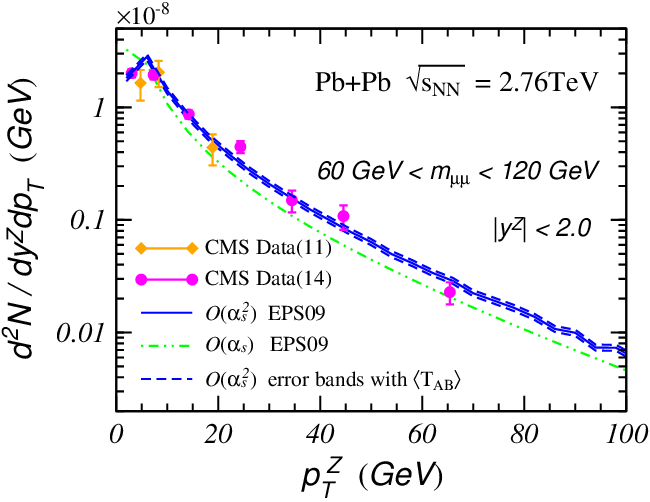}
\caption{(Color online)  The yields $d^2N/dydp_T^Z$ in MB
Pb+Pb collisions at $\sqrt{s_{NN}}=2.76$~TeV. The blue
dash curves are error of the $\mathcal {O}(\alpha_s^2)$
calculations caused by $\langle T_{AB}\rangle_{MB}$.
The CMS data~(muon) are taken from Ref.~\cite{Chatrchyan:2011ua,Chatrchyan:2014csa}. }
\label{fig:Z-ptdata@pbpb}
\end{figure}

Usually to construct a set of nPDFs, we can choose a set of free proton PDFs $f^p(x,\mu)$, and then implement nuclear modifications on them,
where $x$ is the momentum fraction and $\mu$ is the scale.
In parametrizations of nPDFs such as EPS09~\cite{Eskola:2009uj} and DSSZ~\cite{deFlorian:2011fp}, the nuclear modifications to the PDFs are defined through
flavor and scale dependent factors $R_f(x,\mu)$.
One can obtain the parton distribution of the nuclear proton $f^{p,A}(x,\mu)$, by multiplying the factor with the free proton's PDFs as:
\begin{eqnarray}
f^{p,A}(x,\mu)=R_f(x,\mu)f^{p}(x,\mu).
\end{eqnarray}
The PDFs for the nuclear neutron $f^{n,A}(x,\mu)$ can be derived from those of the bound proton by assuming the isospin
symmetry as~\cite{Eskola:2009uj,deFlorian:2011fp},
\begin{eqnarray}
d^{n,A}(x,\mu)&=&u^{p,A}(x,\mu) , \\ \nonumber
u^{n,A}(x,\mu)&=&d^{p,A}(x,\mu).
\end{eqnarray}
The PDFs of other flavor partons in the neutron are the same as those of the proton.
In this way, the nPDFs $f^A(x,\mu)$ can be obtained as ~\cite{Eskola:2009uj,deFlorian:2011fp}
\begin{eqnarray}
f^A(x,\mu) = \frac{Z}{A}f^{p,A}(x,\mu)+\frac{N}{A}f^{n,A}(x,\mu)
\end{eqnarray}
where $Z$ and $N$ are the number of protons and neutrons in a
nucleus with mass number $A$. In this paper, the proton PDFs are taken from MSTW2008 where the NNLO PDFs are provided, and the nuclear PDFs modification factors $R_f(x,\mu)$ are from EPS09 and DSSZ, which are two most widely used parametrization sets of nPDFs.

In the Glauber approach, the cross sections and yields in p+A and A+A collisions, can be derived from the nucleon-nucleon cross section, by
multiplying the number of nucleon-nucleon collisions $\langle N_{\text coll}\rangle$ and the nuclear overlap function $\langle T_{\text AB}\rangle$, respectively~\cite{d'Enterria:2003qs,Loizides:2014vua}.
In this work, the yields of minimal bias (MB) reactions are obtained with the $\langle T_{\text AB}\rangle_{\text MB}$ given by the CMS collaboration.~\cite{Chatrchyan:2011ua}

Recently the $Z^0$ boson productions in p+Pb collisions at $\sqrt{s_{NN}}=5.02$~TeV and in Pb+Pb collisions
at $\sqrt{s_{NN}}=2.76$~TeV have been measured at the LHC~\cite{Chatrchyan:2011ua,Aad:2012ew,atlasz14,CMS:2014sca,Chatrchyan:2014csa}.
As a first example we study
the transverse momentum distributions of $Z^0$ in heavy-ion collisions and the corresponding nuclear modification
factors. Figure~\ref{fig:Z-ptdata@pbpb} shows the theoretical simulations of the yields $d^2N/dydp_T$ of $Z^0\to \mu\mu$ per MB event
as a function of $p_T^Z$ in the $Z^0$ boson rapidity range $|y|<2.0$ for Pb+Pb collisions at $\sqrt{s_{NN}}=2.76$~TeV.
The results are compared with the CMS muon data, which have been measured
in the $Z^0$ invariant mass interval $60$~GeV$<m_{\mu\mu}<120$~GeV~\cite{Chatrchyan:2011ua,Chatrchyan:2014csa}.
Our yields are obtained by multiplying
cross section per nucleon-nucleon collisions by the average nuclear overlap function for
minimum bias collisions $\langle T_{AB}\rangle_{MB}=(5.66\pm0.35)
mb^{-1}$, which is the same as that used in Ref.~\cite{Chatrchyan:2011ua}.
We can see in Fig.~\ref{fig:Z-ptdata@pbpb} that the theory could give a decent description
of the data in Pb+Pb collisions, though at very low $p_T$ region a deviation between theory and data is observed.
We note that at small $p_T$, a good treatment of resummation of leading log or next-to-leading log contributions
in p+p reactions may be needed for a more precise comparison in Pb+Pb collisions.

\begin{figure}[t]
\includegraphics[scale=0.7]{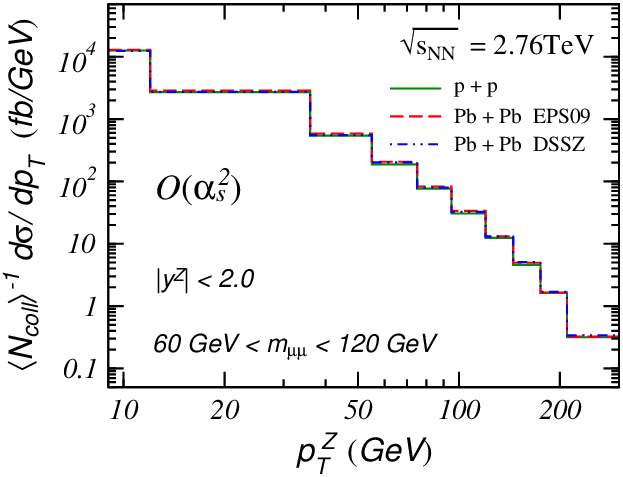}
\caption{(Color online) The differential cross sections
$\langle N_{coll}\rangle^{-1}d\sigma/dp_T^Z$ for the $Z^0$
boson productions in p+p and Pb+Pb collisions at $\sqrt{s_{NN}}=2.76$~TeV. }
\label{fig:Z-pt@pbpb&pp}
\end{figure}

In Fig.~\ref{fig:Z-pt@pbpb&pp} we plot the $p_T^Z$ dependence of the $\langle N_{\text coll}\rangle$ scaled differential cross sections in
Pb+Pb with EPS09 and DSSZ nPDFs as well as that in p+p at $\mathcal {O}(\alpha_s^2)$.
To see the nuclear effects more clearly, we define the nuclear modification ratio $R_{AA}$
as
\begin{eqnarray}
R_{AA}(p_T)=\frac{d\sigma^{\text AA}/dp_T}{\langle N_{\text coll}\rangle d\sigma^{pp}/dp_T} \, \,
\end{eqnarray}
The ratio as functions of other variables, such as rapidity can be defined similarly.

\begin{figure}[t]
\includegraphics[scale=0.7]{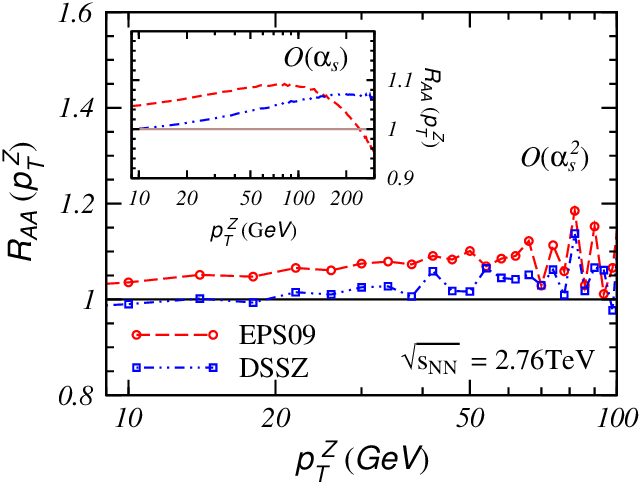}
\caption{(Color online) The $p_T$ dependence of the nuclear modification factors $R_{AA}(p_T)$ for $Z^0$ boson
 at $\mathcal {O}(\alpha_s^2)$. The insert shows the results at  $\mathcal {O}(\alpha_s)$.}
\label{fig:Z-Raapt@pbpb}
\end{figure}

\begin{figure}[t]
\includegraphics[scale=0.7]{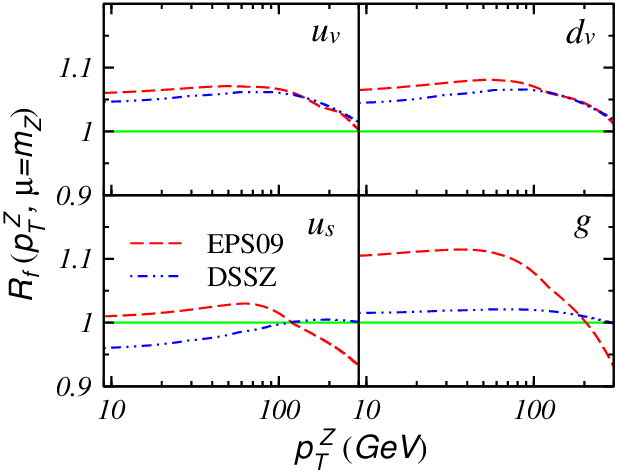}
\caption{(Color online) The flavor dependent nuclear modifications of EPS09
and DSSZ nPDFs at $\sqrt{s_{NN}}=2.76$~TeV.
The factorization scale is fixed at the boson mass.}
\label{fig:Z-Rfpt@pbpb}
\end{figure}

In Fig.~\ref{fig:Z-Raapt@pbpb} the $R_{AA}$ as a function of $p_T^Z$ is plotted.
A small enhancement could be observed in the large $p_T^Z$ region ( $\sim 20 - 200$~GeV)
for both EPS09 and DSSZ nPDFs.
To better understand the $R_{AA}(p_T^Z)$, we made
the following analysis at the order $\mathcal {O}(\alpha_s)$.
At this order, we have the relationship~\cite{He:2011sg} of the momentum fractions $x_1$ and $x_2$ of initial partons with the narrow width approximation assuming $m_{ll}\sim m_{Z/W} $ as
\begin{eqnarray}
x_1&=&(E_{T3}e^{y_3}+E_{T4}e^{y_4})/\sqrt{s},\nonumber  \\
x_2&=&(E_{T3}e^{-y_3}+E_{T4}e^{-y_4})/\sqrt{s},
\end{eqnarray}
where $y_3$ and $y_4$ are the rapidity of the vector boson $V$ and
of the parton recoiling against it, and $E_T$ is the particle transverse energy
defined as $E_T=\sqrt{p_T^2+m^2}$.
At the central rapidity with $y_3\sim y_4\sim 0$,
we derive a simple relationship
\begin{eqnarray}
x_1=x_2=(E_{T3}+E_{T4})/\sqrt{s}.
\end{eqnarray}
In the partonic process,  with
$p_{T3}=p_{T4}$ and by neglecting the parton mass one can obtain in
the narrow width approximation that
\begin{eqnarray}
\label{xpt}
x_1=x_2=\frac{p_T+\sqrt{p_T^2+m_{Z/W}^2}}{\sqrt{s}},
\end{eqnarray}
which connects the initial parton momentum fractions with the gauge boson
transverse momentum in the mid-rapidity region at leading order.

Because of the large mass of $Z^0$ boson~(or $W$ boson),
the momentum fraction $x$ can not be too
small in the mid-rapidity boson production.
Even with $p_T^Z=0$, for heavy-ion collisions at $\sqrt{s_{NN}}=2.76$~TeV,
one obtain $x\sim m_Z/\sqrt{s_{NN}}\sim0.033$, which is
in the vicinity of anti-shadowing region~\cite{Eskola:2009uj,deFlorian:2011fp,Armesto:2006ph} for the EPS09 nPDFs. This underlies
the enhancement at large $p_T^Z$.

\begin{table}[b]
\tabcolsep 0pt \caption{The $Z^0$ boson nuclear modification ratios $R_{AA}$
calculated in $p_T^Z$ bins and the corresponding CMS data~\cite{Chatrchyan:2014csa}.} \vspace*{-12pt}
\begin{center}
\def\temptablewidth{0.45\textwidth}
{\rule{\temptablewidth}{1pt}}
\begin{tabular*}{\temptablewidth}{@{\extracolsep{\fill}}ccccc}
 $R_{AA}(p_T^Z)$ :&   & & &\\
 \hline
&$\mathcal {O}(\alpha_s^2)$ & $\mathcal {O}(\alpha_s^2)$ & & CMS \\
$p_T^Z$(~GeV)& EPS09 & DSSZ & & Data \\
\hline
$[0,5]$ & $0.909$ &$0.901$ & & $0.99\pm0.09\pm0.08$\\
$[5,10]$ & $1.018$ &$0.980$ & & $1.29\pm0.14\pm0.11$\\
$[10,20]$ & $1.071$ &$1.000$ &  & $0.93\pm0.10\pm0.08$ \\
$[20,30]$ & $1.064$ &$1.002$ & & $1.27\pm0.20\pm0.11$\\
$[30,40]$ & $1.028$ &$1.008$ & & $1.18\pm0.31\pm0.10$\\
$[40,50]$ & $1.129$ &$1.068$ &  & $1.28\pm0.40\pm0.11$ \\
$[50,100]$ & $1.084$ &$1.049$ &  & $0.89\pm0.28\pm0.07$ \\
 \hline
\end{tabular*}
{\rule{\temptablewidth}{1pt}}
\end{center}
\label{table:Z-Raaptdata@pbpb}
\end{table}

Figure~\ref{fig:Z-Raapt@pbpb} shows that differences exist between the results of EPS09 and DSSZ nPDFs.
For example, more enhancement is
given by EPS09 in the region of $p_T^Z\lesssim130$~GeV, but DSSZ gives persistent
enhancement even when the yields of EPS09 go down in the larger $p_T^Z$ region
( $> 130$~GeV), which can be seen more obviously in the insert of Fig.~\ref{fig:Z-Raapt@pbpb}.
To have a deeper understanding of these differences, we plot in Fig.~\ref{fig:Z-Rfpt@pbpb} the flavor dependent factors of both
EPS09 and DSSZ  as $R_f(p_T^Z,\mu)$ by replacing the parton momentum fraction $x$
with $p_T^Z$, where the relation of $x$ and $p_T^Z$ is given by Eq.~(\ref{xpt}).
One can observe that the nuclear modification factors of the $u$ and $d$ valence quarks
are similar for EPS09 and DSSZ. However, obvious distinctions could be found for
$u$ sea quarks~(those for $d$ sea quarks are similar as $u$ sea quarks, thus aren't shown), especially for gluons.
In the region of $p_T^Z\lesssim130$~GeV, the nuclear PDF of
$u$ sea quark is shadowed in DSSZ while it is anti-shadowed in EPS09, and the anti-shadowing
effects for gluons are very weak in DSSZ but much stronger in EPS09. Because partonic subprocesses with at least one initial-state gluon (for example, $u_v+g\rightarrow u+Z^0$ and $d_v+g\rightarrow d+Z^0$ at LO) give dominant contributions to $Z^0$ production
at large $p_T$,  $R_{AA}(p_T^Z)$ could provide a good tool to distinguish different parametrizations of gluon distributions in nuclei~\cite{Kang:2012am}.
In TABLE ~\ref{table:Z-Raaptdata@pbpb} we have listed the theoretical results and CMS data for $R_{AA}$. It could be seen that the numerical results of $R_{AA}$ in both EPS09 and DSSZ show a suppression at small $p_T$ and an enhancement at large $p_T$, whereas the CMS data give a zigzag
dependence on $p_T$. However, because the experimental uncertainty of the CMS data is rather large, the theoretical results still roughly fall inside the experimental band with the large experimental error bars, and a more precise measurement will be needed for a robust comparison between the theory and the data.

\begin{figure}[t]
\includegraphics[scale=0.7]{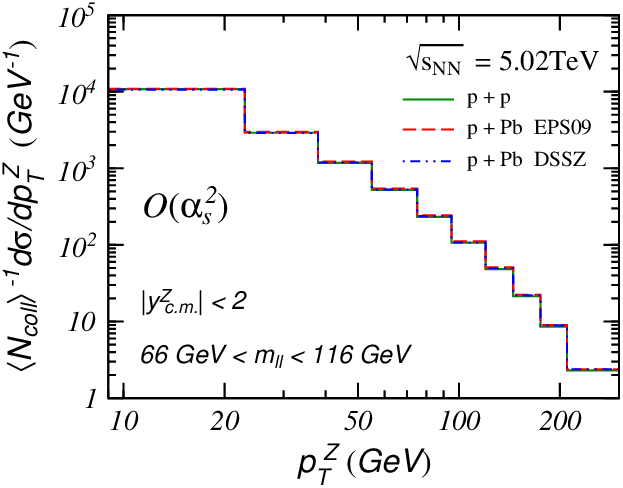}
\caption{(Color online) Differential cross section for the $Z^0$ boson production
in p+Pb collisions at $\sqrt{s_{NN}}=5.02$~TeV.}
\label{fig:Z-pt@ppb&pp}
\end{figure}

For p+Pb collisions at $\sqrt{s_{NN}}=5.02$~TeV, the differential
cross sections as a function of $p_T^Z$
are calculated in the $Z^0$ boson invariant
mass interval $66$~GeV$<m_{ll}<116$~GeV, as adapted by the ATLAS experiment~\cite{Aad:2011gj,atlasz14}.
What should be mentioned for p+Pb collisions is that for the asymmetric collision system,
the center of mass frame for the colliding nucleon pair is not
the same as the laboratory frame. At the LHC, the relationship of rapidity between them is
$y_{c.m.}=y_{lab}-0.465$. Here we calculate the $p_T^Z$ distribution in the rapidity region
$|y_{c.m.}|<2$.
In Fig.~\ref{fig:Z-pt@ppb&pp} we show results with EPS and DSSZ nPDFs at the $\mathcal {O}(\alpha_s^2)$ and find
the difference is small.

\begin{figure}[t]
\includegraphics[scale=0.7]{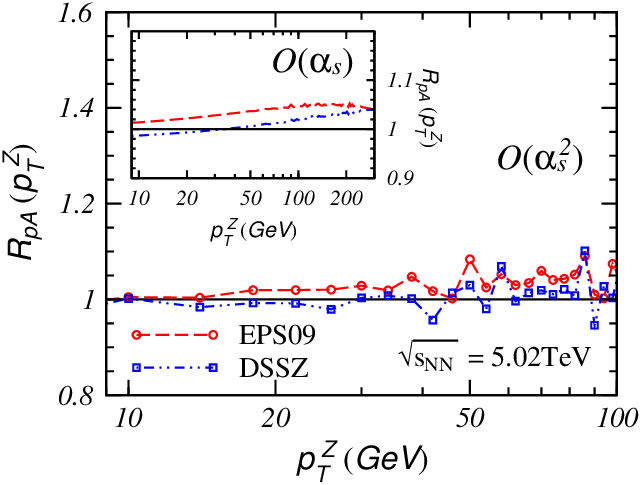}
\caption{(Color online) The nuclear modification factors $R_{pA}$ for the $Z^0$ production as a function of $p_T^Z$
at $\mathcal {O}(\alpha_s^2)$. The insert shows the results at $\mathcal {O}(\alpha_s)$.}
\label{fig:Z-Rpapt@ppb}
\end{figure}

The nuclear modification ratios $R_{pA}(p_T^Z)$
are plotted in Fig.~\ref{fig:Z-Rpapt@ppb}. For EPS09 nPDFs, enhancements could be
observed in the $p_T^Z$ region of about $10\sim300$~GeV. For DSSZ, suppressions exist in
the region $p_T^Z\lesssim40$~GeV and enhancements appear in the larger $p_T^Z$ region.
Compared to results in Pb+Pb, the
suppression and enhancement regions in p+Pb become wider because of the larger colliding
energy $\sqrt{s_{NN}}$ in p+Pb. For example, with $p_T=200$~GeV
the Eq.~(\ref{xpt}) gives the parton momentum fraction $x\sim0.083$ for p+Pb in the central rapidity $y_3\sim y_4\sim 0$, which is
in the anti-shadowing region in EPS09. But the momentum fraction $x$ given by Eq.~(\ref{xpt}) may fall in the EMC region~\cite{Eskola:2009uj,deFlorian:2011fp,Armesto:2006ph}
for Pb+Pb collisions at $2.76$~TeV. Also we note the CNM effects will be less pronounced for p+Pb, the smaller colliding system
as compared to those in Pb+Pb.

\begin{figure}[t]
\includegraphics[scale=0.7]{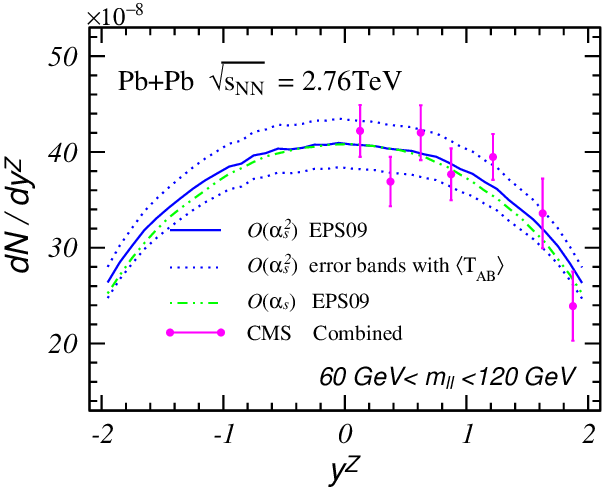}
\caption{(Color online)  The yields
$dN/dy$  versus the $Z^0$ boson rapidity $y^Z$ in MB Pb+Pb collisions. The blue dotted dash
curves represent errors of the $\mathcal {O}(\alpha_s^2)$ calculations
caused by $\langle T_{AB}\rangle_{MB}$. The data are taken from CMS~\cite{Chatrchyan:2014csa}.}
\label{fig:Z-ydata@pbpb}
\end{figure}

Next we focus on another important observable in $Z^0$ boson productions, i.e. its rapidity distribution.
The yields per MB event $dN/dy$ in Pb+Pb collisions at $\sqrt{s_{NN}}=2.76$~TeV
are calculated numerically and compared to the CMS data~\cite{Chatrchyan:2011ua,Chatrchyan:2014csa} in Fig.~\ref{fig:Z-ydata@pbpb}.
One can observe that results at $\mathcal {O}(\alpha_s^2)$ and
$\mathcal {O}(\alpha_s)$ are close to each other, and
$\mathcal {O}(\alpha_s^2)$ calculations slightly enhance the total yields.
We also note that the experimental error bars are rather large and to make a robust comparison more precise measurement may be needed.

\begin{figure}[t]
\includegraphics[scale=0.7]{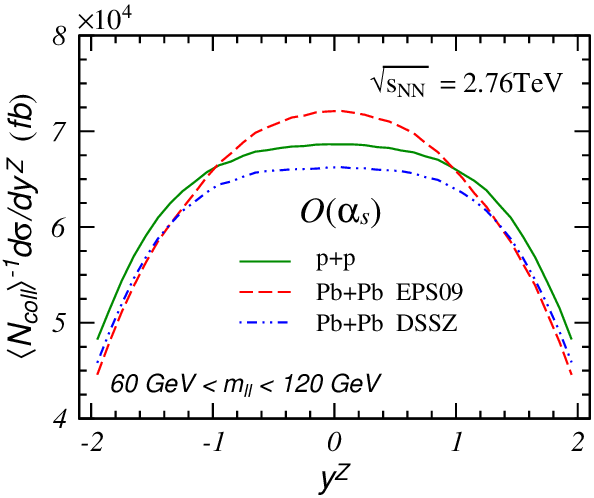}
\caption{(Color online)  The $Z^0$ boson differential cross section  in p+p and Pb+Pb collisions
 at $\mathcal {O}(\alpha_s)$ as a function of rapidity. }
\label{fig:Z-y@pbpb}
\end{figure}

\begin{figure}[t]
\includegraphics[scale=0.7]{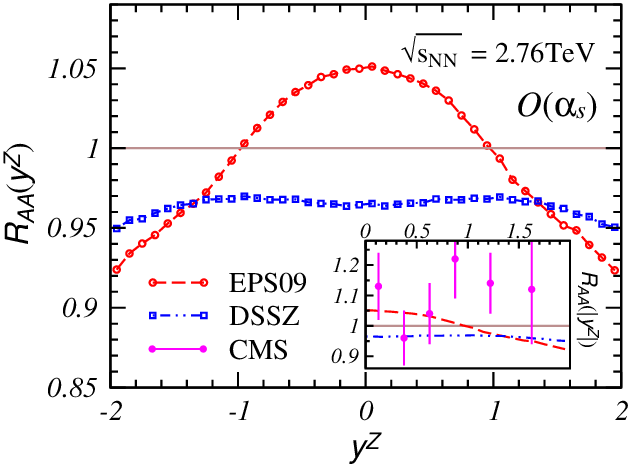}
\caption{(Color online) the nuclear modification factors $R_{AA}(y^Z)$ for the $Z^0$ boson production
at $\mathcal {O}(\alpha_s)$. The CMS data are taken from Ref.~\cite{Chatrchyan:2014csa}. }
\label{fig:Z-Raay@pbpb}
\end{figure}

In Fig.~\ref{fig:Z-y@pbpb} we compare the differential cross section of $Z^0$ in Pb+Pb with that in p+p,
and give the related nuclear modification ratios $R_{AA}(y^Z)$ in Fig.~\ref{fig:Z-Raay@pbpb}.
In Pb+Pb collisions, considerable distinction between results with EPS09 nPDFs and those with DSSZ nPDFs could be observed.
Relative to that in p+p, the yield of Pb+Pb with EPS09 shows an enhancement in the mid-rapidity region
(for $|y^Z|<1$), but a suppression in the large rapidity region (for $1<|y^Z|<2$).
However, the yield with DSSZ always gives a considerable suppression in the whole rapidity region we studied.
For convenience, we also insert in Fig.~\ref{fig:Z-Raay@pbpb} a comparison between the theoretical results and the CMS data on $R_{AA}(|y^Z|)$.

\begin{figure}[t]
\includegraphics[scale=0.7]{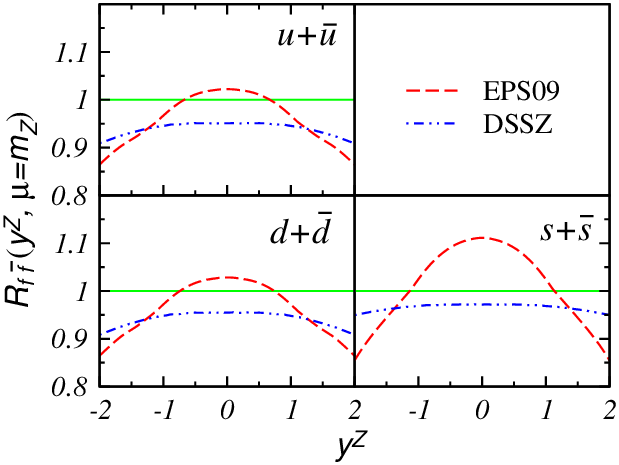}
\caption{(Color online) The flavor dependent factor $R_{f\bar{f}}(y^Z,\mu)$ as
functions of $Z^0$ boson rapidity. The factorization scale is fixed at the boson mass.}
\label{fig:Z-Rffbary@pbpb}
\end{figure}

To have a simple picture of the distinction between
results with EPS09 and DSSZ, one can look at the leading-order processes for $Z^0$ production.
At LO, the $Z^0$ boson rapidity can be connected to the parton (quark and anti-quark)
momentum fractions with the kinematic relations
\begin{eqnarray}
\label{xy}
x_1=\frac{m_Z}{\sqrt{s_{NN}}}e^{y^Z},
x_2=\frac{m_Z}{\sqrt{s_{NN}}}e^{-y^Z}.
\end{eqnarray}
Then at $y^Z=0$ one obtains $x_1=x_2=m_Z/\sqrt{s_{NN}}\sim0.033$,
at which the anti-shadowing effect is nearly peaked in EPS09 parametrization. Therefore
the visible enhancement of $Z^0$ yield with EPS09 in Pb+Pb relative to that in p+p at $y^Z=0$ originates mainly from the anti-shadowing
effects. When going to the large $|y^Z|$ region, $Z^0$ bosons are produced
with one of parton momentum fractions ($x_1$ or $x_2$) increasing and going to
the EMC effect region, while the other decreasing and going to the shadowing region~\cite{Eskola:2009uj,deFlorian:2011fp,Armesto:2006ph}.
As a consequence, a suppression gradually developes in the large $|y^Z|$ regime.
For example, if we consider $Z^0$ boson production at $y^Z=2$, then one gets $x_1\sim0.24$ and $x_2\sim0.0045$.
For EPS09 nPDFs, $x_1$ has entered the EMC region, and $x_2$ has entered the
shadowing region, so a suppression is observed at $y^Z=2$ since both shadowing and EMC effects
may reduce the total yield of $Z^0$ gauge bosons.

~

To demonstrate the differences between
EPS09 and DSSZ, we plot in Fig.~\ref{fig:Z-Rffbary@pbpb} the factor $R_{f\bar{f}}(y^Z,\mu)$ for LO subprocess $f+\bar{f}\rightarrow Z^0$ defined as
\begin{eqnarray}
\label{rffbar}
R_{f\bar{f}}(y^Z,\mu)=&\frac{1}{2}&[R_f(x_1,\mu)R_{\bar{f}}(x_2,\mu) \nonumber \\
&+&R_f(x_2,\mu)R_{\bar{f}}(x_1,\mu)] \,\, ,
\end{eqnarray}
where $x_{1,2}$ is related to $Z^0$ rapidity according to Eq.~(\ref{xy}).
By comparing Fig.~\ref{fig:Z-Rffbary@pbpb} with Fig.~\ref{fig:Z-Raay@pbpb}, one can observe that the factor $R_{f\bar{f}}$ roughly reflects the
nuclear modification of $Z^0$ production.
One can also see clearly in Fig.~\ref{fig:Z-Rffbary@pbpb} that the three partonic subprocesses
($u+\bar u \rightarrow Z^0$, $d+\bar d \rightarrow Z^0$, $s+\bar s \rightarrow Z^0$) suffer similar
nuclear modifications.  They demonstrate that $R_{AA}(y^Z)$ results mainly from the nuclear modifications
of quark and anti-quark distributions, and thus provide an excellent observable to measure the nuclear
modification of (anti-)quark distribution.  On the contrary as we have discussed
$R_{AA}(p_T^Z)$ (or $R_{pA}(p_T^Z)$) gives more information about the nuclear effects on gluons.

\begin{figure}[t]
\includegraphics[scale=0.7]{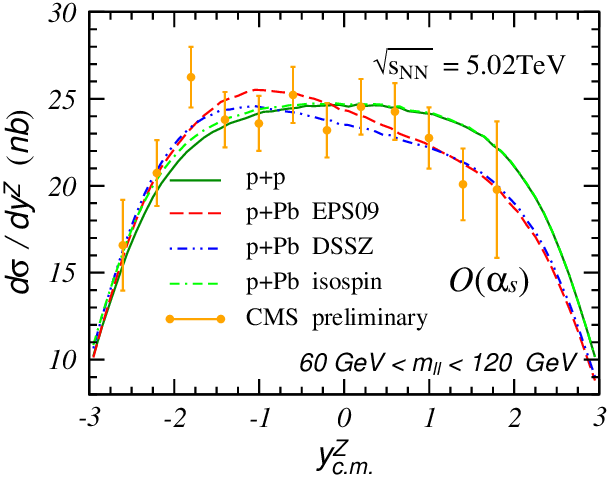}
\caption{(Color online) The differential cross section
for $Z^0$ boson  in p+Pb and p+p collisions.
The green dotted-dash line stands for the result by only considering the isospin effect
in p+Pb collisions. The CMS preliminary data are taken from Ref.~\cite{CMS:2014sca}.}
\label{fig:Z-y@ppb&pp}
\end{figure}

\begin{figure}[t]
\includegraphics[scale=0.7]{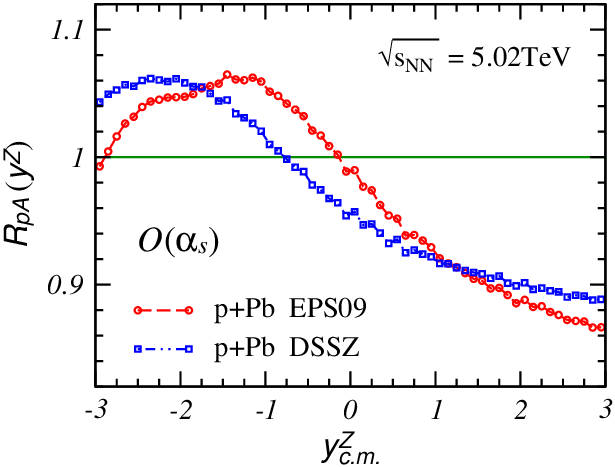}
\caption{(Color online) the nuclear modification factors $R_{pA}(y^Z)$ for the $Z^0$ production
 at $\mathcal {O}(\alpha_s)$.}
\label{fig:Z-Rpay@ppb}
\end{figure}

\begin{figure}[t]
\includegraphics[scale=0.7]{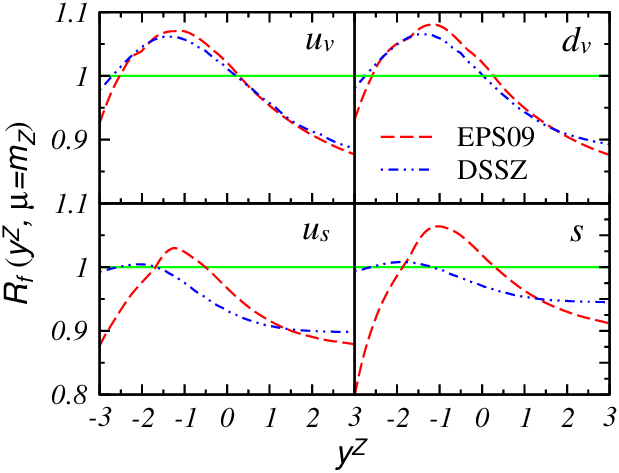}
\caption{(Color online) The flavor dependent nuclear modification factor of EPS09
and DSSZ nPDFs as a function of $y^Z$. The factorization scale is fixed at the boson mass.}
\label{fig:Z-Rfy@ppb}
\end{figure}

We also investigate the rapidity distribution of $Z^0$ in p+Pb collisions at $\sqrt{s_{NN}}=5.02$~TeV,
and the results are shown in Fig.~\ref{fig:Z-y@ppb&pp}. $Z^0$ bosons are produced in the CMS mass window
$60$~GeV$<m_{ll}<120$~GeV~\cite{CMS:2014sca}. Figure~\ref{fig:Z-y@ppb&pp} shows
a forward-backward asymmetry of $Z^0$ boson rapidity
distributions in p+Pb collisions. Compared to p+p at $\sqrt{s}=5.02$~TeV,
a suppression exists in the forward rapidity region and the enhancement shows up
in the backward direction in p+Pb collisions. We have checked that the isospin effect
just gives a slight contribution to the nuclear modifications observed in $Z^0$ rapidity
distribution, which implies the forward-backward asymmetry results mainly from other CNM effects,
such as the shadowing effect. In Fig.~\ref{fig:Z-Rpay@ppb}
the nuclear modification factors $R_{pA}(y^Z)$ are given.
Using Eq.~(\ref{xy}) one finds at $y^Z=0$, the momentum fractions of partons
$x_1=x_2=m_Z/\sqrt{s_{NN}}\sim0.018$, which is near the junction of the
shadowing and anti-shadowing region of EPS09 nuclear modifications.
Thus at $y^Z=0$ one finds the nuclear effects of $Z^0$ boson with EPS09 are rather moderate.
Moreover, when moving to the forward rapidity regime, parton distributions of the lead nucleus at smaller $x$  are
depleted as compared to those in proton, we expect to get a suppression due to the shadowing effects. And
when going to the backward regime, an enhancement occurs due to the anti-shadowing
effects with the larger $x$ carried by partons in the lead nucleus.
We also show the flavor dependent nuclear modification factor
$R_f(y^Z,\mu)$ in Fig.~\ref{fig:Z-Rfy@ppb} by replacing the parton momentum fraction
$x_2$(from the lead nucleon) with the boson rapidity $y^Z$ by using the Eq.~(\ref{xy}).
One can see that although nPDFs of valence quarks are essentially the same in EPS09 and in DSSZ,
visible difference could be observed for nPDFs of sea quark distribution.

\section{$W^{\pm}$ boson production in heavy-ion collisions}
\label{section:W@HIC}
In Section \ref{section:Z@HIC} we discussed $Z^0$ production in nuclear collisions, with the same approach
we can also investigate $W^{\pm}$ gauge boson production in heavy-ion collisions at NLO and NNLO
with DYNNLO.

\begin{figure}[t]
\includegraphics[scale=0.7]{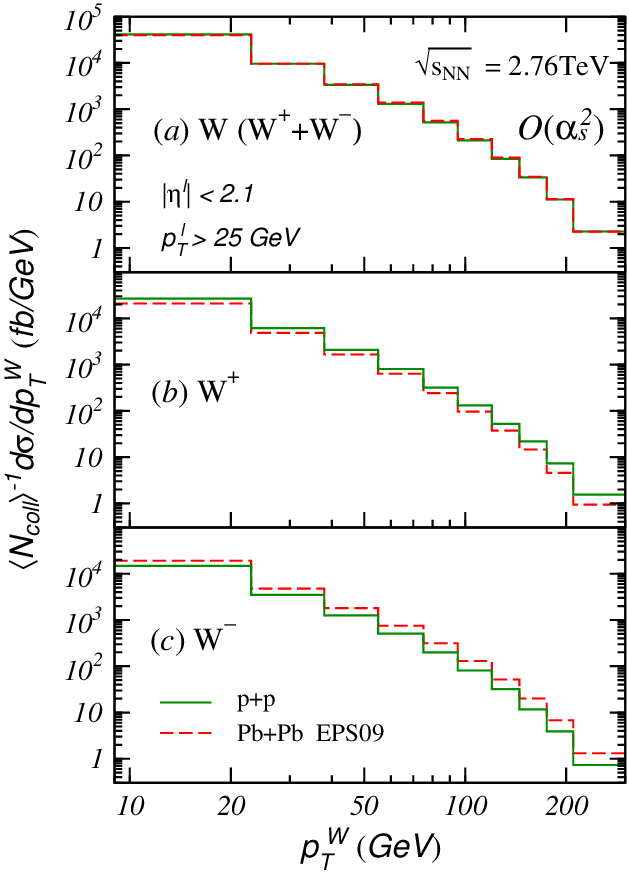}
\caption{(Color online) The differential cross section for $W$ production as a function of $W$ boson transverse
momentum $p_T^W$ in p+p and Pb+Pb collisions with $\sqrt{s_{NN}}=2.76$~TeV at $\mathcal {O}(\alpha_s^2)$. }
\label{fig:W-pt@pbpb&pp}
\end{figure}

First, we study the transverse momentum distribution of
$W$ bosons production. We plot in Fig.~\ref{fig:W-pt@pbpb&pp} the $p_T^W$ differential cross sections
calculated at $\mathcal{O}(\alpha_s^2)$ order.
The results of $W^+$, $W^-$ and $W$($=W^++W^-$) in both p+p and Pb+Pb collisions at
$\sqrt{s_{NN}}=2.76$~TeV are shown. The final state phase space is chosen to be the same as that in the
CMS experiment, which is defined by the transverse momentum and
pseudo-rapidity of the charge lepton (muon at CMS) as $p_T^{l}>25$~GeV and
$|\eta^{l}|<2.1$~\cite{Chatrchyan:2012nt}.
One can observe that, as at $\sqrt{s}=7$~TeV (see Fig.~\ref{fig:W-pt@pp}),
the yields of $W^+$ and $W^-$ in p+p collisions at $\sqrt{s}=2.76$~TeV are quite different,
which is a unique observable in $W^+/W^-$ production and is referred as the charge asymmetry
in the W productions.

The $W^+/W^-$ charge asymmetry originates from the asymmetry in proton
parton distributions. A dominant contribution to this asymmetry comes from the
difference between parton distributions  of $u$ quark and $d$ quark
in a proton, and a moderate contribution from the small deviation between PDFs
of $\bar{u}$ and $\bar{d}$. When computing the transverse momentum spectra of the $W^+/W^-$ productions , the LO processes such as
$u+\bar{d}\to W^++g$ and $d+\bar{u}\to W^-+g$ connect
$W^+$ productions with $u$ and $\bar{d}$, and $W^-$ with $d$ and $\bar{u}$. While
other gluon initiated LO processes like $u+g\to W^++d$ and $d+g\to W^-+u$ also bring considerable contributions to the asymmetry of
the $W^+/W^-$ productions. Therefore the measurement of $W$ boson charge asymmetry might give very important
constraints on the quark parton distribution functions(PDFs), especially the
ratio $u(x)/d(x)$ or $\bar{u}(x)/\bar{d}(x) $~\cite{Yang:2009hh}.

However, in Pb+Pb collisions the yield of $W^+$ is suppressed significantly as compared to that in p+p reactions, whereas
the yield of $W^-$ productions is strongly enhanced.
The opposite trends of $W^+$ and $W^-$ modifications in Pb+Pb mainly result from the
isospin effect since the existence of neutrons in the large nucleus should
significantly increase (decrease)  nuclear parton distribution of $d$ quark ($u$ quark).
We emphasize that although the differential cross sections of $W^+$ and $W^-$ in Pb+Pb are modified significantly relative to
those in p+p, the alteration of the yield for total $W$ production in Pb+Pb is rather small.

\begin{figure}[t]
\includegraphics[scale=0.7]{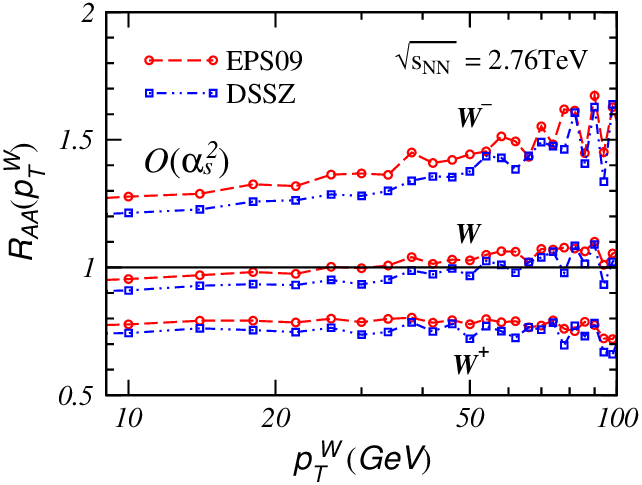}
\caption{(Color online) the nuclear modification factors $R_{AA}(p_T)$ for the $W$ production
at $\mathcal {O}(\alpha_s^2)$. }
\label{fig:W-Raapt@pbpb}
\end{figure}

In Fig.~\ref{fig:W-Raapt@pbpb} we plot nuclear modification ratio $R_{AA}(p_T^W)$ at
$\mathcal {O}(\alpha_s^2)$. As we have discussed
the isospin effect gives the dominant contribution to the distinct behavior of
$R_{AA}^{W^+}$ and $R_{AA}^{W^-}$. It is observed that the enhancement of $W^-$
production becomes stronger with increasing $p_T^W$, and the considerable suppression of $W^+$ is persistent, which results mainly from the
increased $u-d$ asymmetry of PDFs at large momentum fraction $x$.

To better understand the isospin effect and the obvious separation
of $R_{AA}^{W^+}$ and $R_{AA}^{W^-}$, we define
the  asymmetry ratio of parton PDFs in a proton as
\begin{eqnarray}
\label{rud}
r_{ud}(x)=\frac{u(x)-d(x)}{u(x)+d(x)},\nonumber \\
r_{\bar{u}\bar{d}}(x)=\frac{\bar{d}(x)-\bar{u}(x)}{\bar{d}(x)+\bar{u}(x)}\, .
\end{eqnarray}
At LO the $W$ bosons are mainly produced through partonic subprocess with an initial-state gluon,
such as $u+g\rightarrow W^++d$ and $\bar{d}+g\rightarrow W^++\bar{u}$ for $W^+$ production,
and $d+g\rightarrow W^-+u$ and $\bar{u}+g\rightarrow W^-+\bar{d}$ for $W^-$ production.
If only considering the isospin effect and neglecting the contributions of other nuclear effects~(included in the factor $R_f(x)$),
one can write the nuclear parton distribution functions as
\begin{eqnarray}
f^A(x) \approx \frac{Z}{A}f^{p}(x)+\frac{N}{A}f^{n}(x).
\end{eqnarray}
Then we obtain some simple relations as:
\begin{eqnarray}
\label{raarud}
\frac{u^{A}(x)}{u^{p}(x)}&\approx&\frac{Z-N}{A}+\frac{2N}{A}\frac{1}{1+r_{ud}(x)} \nonumber\\
\frac{d^{A}(x)}{d^{p}(x)}&\approx&\frac{Z-N}{A}+\frac{2N}{A}\frac{1}{1-r_{ud}(x)},
\end{eqnarray}
where $f^A(x)$ is the parton distribution in nuclei and $f^p(x)$ is that in proton. More details can be found in the Appendix~\ref{section:appendix}.
In Fig.~\ref{fig:W-Raaptiso@pbpb} we plot  $f^A(p_T^W)/f^p(p_T^W)$ by replacing the parton momentum
fraction $x$ with $p_T^W$ according to Eq.~(\ref{xpt}). By comparing Fig.~\ref{fig:W-Raapt@pbpb} with
Fig.~\ref{fig:W-Raaptiso@pbpb} we can see clearly that the isospin effect gives the most important
contribution to $R_{AA}^{W^+}(p_T)$ and $R_{AA}^{W^-}(p_T)$.

Certainly other CNM effects, especially the (anti-)shadowing may also affect the $R_{AA}(p_T^W)$ of
$W^+$, $W^-$ and total $W$. In Fig.~\ref{fig:W-Raapt@pbpb}
we give both the results with EPS09 and those with DSSZ nPDFs. Similar differences between the two nPDFs for $W^+$, $W^-$ and total $W$
could be observed, which is similar as we learned in the $Z^0$ production.

\begin{figure}[t]
\includegraphics[scale=0.7]{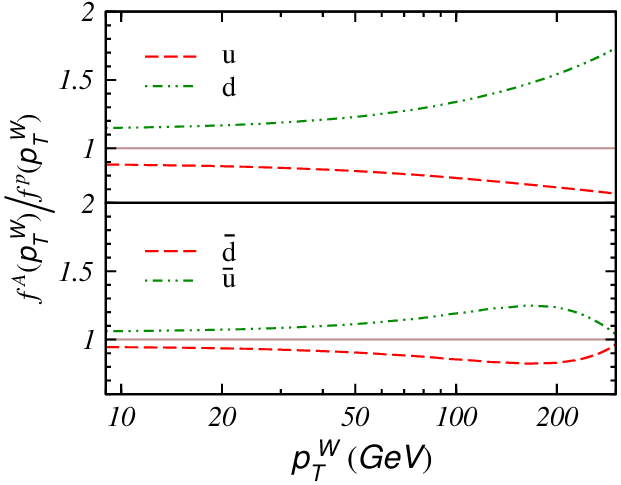}
\caption{(Color online) The flavor dependent nuclear isospin effect factor $f^A(p_T^W)/f^p(p_T^W)$
given by MSTW2008 NLO PDFs, and the factorization scale is fixed at the boson mass.}
\label{fig:W-Raaptiso@pbpb}
\end{figure}

\begin{figure}[t]
\includegraphics[scale=0.7]{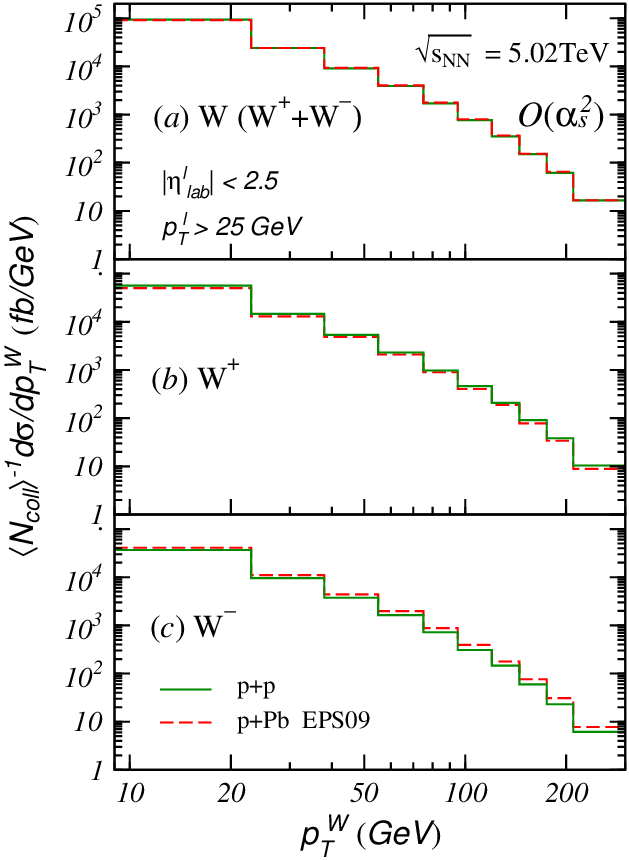}
\caption{(Color online) The differential cross section
 for the $W$ bosons production in p+p and p+Pb collisions at $\mathcal {O}(\alpha_s^2)$ at $\sqrt{s_{NN}}=5.02$~TeV.}
\label{fig:W-pt@ppb&pp}
\end{figure}

\begin{figure}[t]
\includegraphics[scale=0.7]{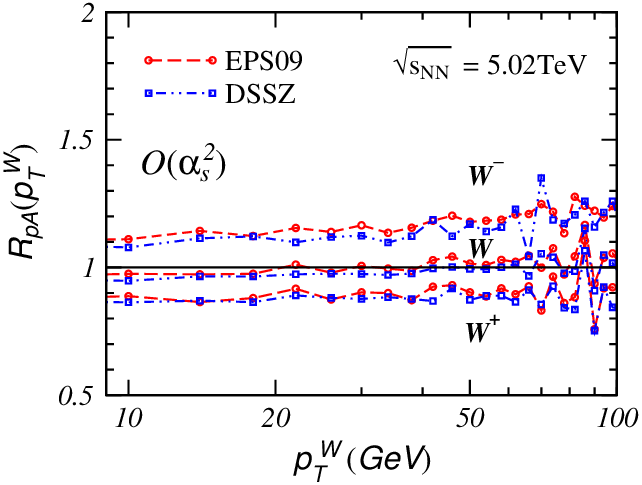}
\caption{(Color online) The nuclear modification ratio $R_{pA}(p_T^W)$ for the $W$  production
at $\mathcal{O}(\alpha_s^2)$. }
\label{fig:W-Rpapt@ppb}
\end{figure}
To complement the study of the $W$ production in Pb+Pb, we also carry out the numerical simulations
of the $W$ boson production in p+Pb collisions at $\sqrt{s_{NN}}=5.02$~TeV. The $p_T^W$ differential cross sections
are shown in Fig.~\ref{fig:W-pt@ppb&pp} and the nuclear modification factors $R_{pA}(p_T^W)$ in Fig.~\ref{fig:W-Rpapt@ppb}.
The calculations are made in the CMS final-state phase space defined as $p_T^{l}>25$~GeV and
$|\eta^{l}_{lab}|<2.5$~\cite{CMS:2014kla}. Compared to that in Pb+Pb collisions, the isospin effect results in a similar but weaker
nuclear modification in p+Pb collisions.

To study the difference of $W^+$ and $W^-$ in p+p and heavy-ion collisions, CMS collaboration has measured the so-called
charge asymmetry observable defined as
\begin{eqnarray}
A=\frac{N_{W^+}-N_{W^-}}{N_{W^+}+N_{W^-}}\,\,\, .
\end{eqnarray}

\begin{figure}[t]
\centering
\includegraphics[scale=0.7]{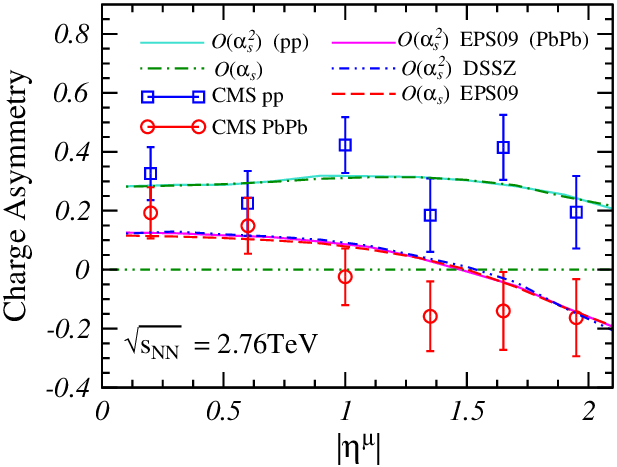}
\caption{(Color online) Charge asymmetry $ (N_{W^+}-N_{W^-})/(N_{W^+}+N_{W^-})$
as a function of charged lepton pseudorapidity in p+p and Pb+Pb at $\sqrt{s_{NN}}=2.76$~TeV.
The CMS data are taken from Ref.~\cite{Chatrchyan:2012nt}}
\label{fig:W-cas@pbpb&pp}
\end{figure}

Figure~\ref{fig:W-cas@pbpb&pp} shows the dependence of the charge asymmetry observable on pseudorapidity of the absolute value of
charged lepton at $\mathcal {O}(\alpha_s)$ (NLO) and $\mathcal{O}(\alpha_s^2)$ (NNLO) for both p+p and Pb+Pb collisions.
One could observe that for p+p reactions,
the charge asymmetry is always positive in the whole region
$|\eta^{\mu}|<2.1$. For Pb+Pb, the charge asymmetry is
positive in $|\eta^{\mu}|\lesssim 1$ and then decreases to be
negative in $1\lesssim |\eta^{\mu}|\lesssim 2.1$. And
the NNLO calculations give negligible corrections to the NLO results.
It is also found that EPS09 and DSSZ give very similar charge asymmetries in Pb+Pb collisions,
even though the two nPDFs give different nuclear modifications $R_f(x,\mu)$.
Our theoretical calculation give a decent description of the CMS data~\cite{Chatrchyan:2012nt}.

\begin{figure}[h]
\includegraphics[scale=0.7]{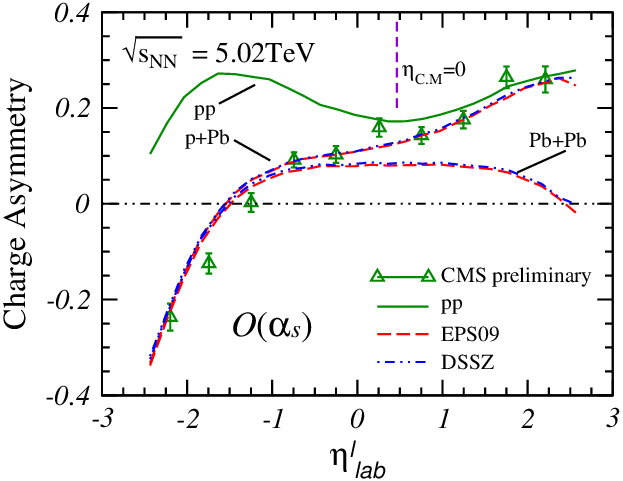}
\caption{(Color online) The charge asymmetry $(N_{W^+}-N_{W^-})/(N_{W^+}+N_{W^-})$ as a function of
the charged lepton pseudorapidity for $W$ boson productions in p+p, p+Pb and Pb+Pb at $\sqrt{s_{NN}}=5.02$~TeV.
The CMS preliminary data for p+Pb collisions are taken from Ref.~\cite{CMS:2014kla}.}
\label{fig:W-cas@ppb&pbpb&pp}
\end{figure}

In Fig.~\ref{fig:W-cas@ppb&pbpb&pp} we provide the numerical results on charge asymmetry as a function of the charged
lepton pseudorapidity in p+A collisions at $\sqrt{s_{NN}}=5.02$~TeV.
The results agree well with the CMS preliminary data~\cite{CMS:2014kla}.
The theoretical simulation of the charge asymmetry observable in p+p and Pb+Pb collisions at $5.02$~TeV are also
plotted for comparison, which are symmetric with $\eta_{c.m.}^l$, the charged
lepton pseudorapidity in the center-of-mass frame.

It is interesting to observe that the curve of p+Pb collisions is lying between
those of p+p and Pb+Pb collisions.
To be specific, it goes gradually closer to the p+p curve in the forward direction,
while it approaches the Pb+Pb curve in the backward direction.
As in our previous discussions, the differences among these curves mainly originate
from the isospin effect.
And the magnitude of the isospin effect is related
to parton distribution asymmetry ratios such as $r_{ud}(x)$.
In p+Pb collisions, the $W$ production at very forward region is dominated by sub-processes with nuclear partons
at very small momentum fraction $x$, where $r_{ud}(x)$ (or $r_{ \bar{u}\bar{d} }(x)$ ) is rather small. So the isospin effect
becomes weaker in that region. With similar arguments we can find that
the charge asymmetry in the very backward region for Pb+Pb collisions will be very close to that for p+Pb collisions.

\begin{figure}[t]
\includegraphics[scale=0.7]{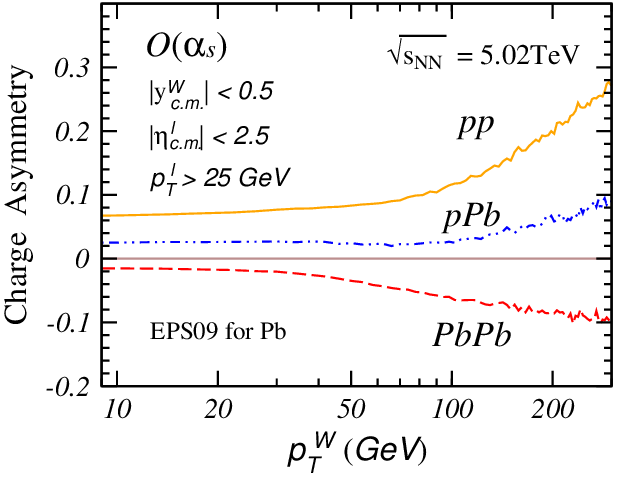}
\caption{(Color online) The charge asymmetry $(N_{W^+}-N_{W^-})/(N_{W^+}+N_{W^-})$ as a function of
the transverse momentum $p_T^W$ in p+p, p+Pb and Pb+Pb at $\sqrt{s_{NN}}=5.02$~TeV.}
\label{fig:W-caspt@ppb&pbpb&pp}
\end{figure}

\begin{table}[b]
\tabcolsep 0pt \caption{The inclusive cross sections for
$W^+$, $W^-$ and total $W$ at $\sqrt{s_{NN}}=2.76$~TeV, and
the corresponding nuclear modification factor $R_{AA}$.
The CMS data are taken from Ref.~\cite{Chatrchyan:2012nt}} \vspace*{-12pt}
\begin{center}
\def\temptablewidth{0.45\textwidth}
{\rule{\temptablewidth}{1pt}}
\begin{tabular*}{\temptablewidth}{@{\extracolsep{\fill}}cccc}
{\rm W production} & $\mathcal {O}(\alpha_s^2)$  & $\mathcal {O}(\alpha_s^2)$ & CMS norm.
cross
\\{\rm Process
} &EPS09 & DSSZ  & section [nb]/$\Delta\eta$ \\
\hline
 $PbPb\to W^+$ & $0.235$ & 0.229 & $0.28\pm0.02\pm0.02$ \\
$PbPb\to W^-$ & $0.212$ & 0.205 & $0.27\pm0.02\pm0.02$\\
\hline
$pp\to W^+$ & $0.316$ & ($\leftarrow$MSTW) & $0.34\pm0.02\pm0.02$\\
$pp\to W^-$ & $0.178$ & & $0.18\pm0.01\pm0.01$\\
\hline
\hline
& $\mathcal {O}(\alpha_s^2)$ & $\mathcal {O}(\alpha_s^2)$ & CMS \\
$R_{AA}$& EPS09 & DSSZ  & Data \\
\hline
$W^+$ & $0.744$ & 0.725 & $0.82\pm0.07\pm0.09$\\
$ W^-$ & $1.191$ & 1.152 & $1.46\pm0.14\pm0.16$\\
$W$ & $0.905$ & 0.879 & $1.04\pm0.07\pm0.12$ \\
 \hline
\end{tabular*}
{\rule{\temptablewidth}{1pt}}
\end{center}
\label{table:W-yield&Raa@pbpb}
\end{table}

\begin{figure}[t]
\includegraphics[scale=0.7]{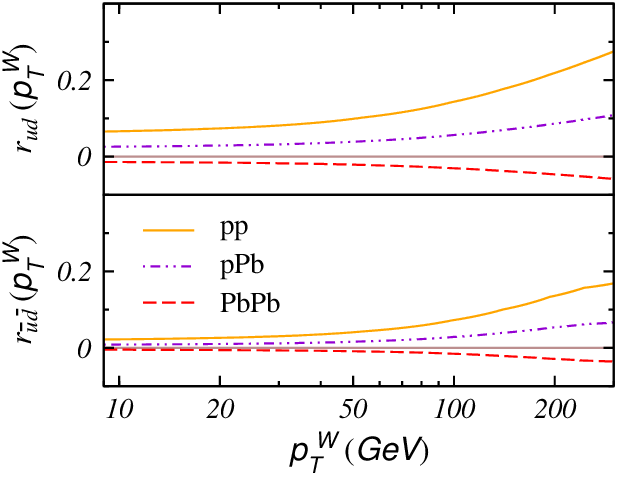}
\caption{(Color online) The parton asymmetry ratio $r_{ud}$ and $r_{\bar{u}\bar{d}}$
as functions of the boson transverse momentum $p_T^W$ in p+p, p+Pb and Pb+Pb at $\sqrt{s_{NN}}=5.02$~TeV.
The results are given by MSTW2008 PDFs at NLO,
and the factorization scale is fixed at the boson mass.}
\label{fig:W-casptiso@ppb&pbpb&pp}
\end{figure}

Only the pseudorapidity dependence of the charge asymmetry was measured so far at the LHC.~\cite{Chatrchyan:2012nt,CMS:2014kla,Aad:2014bha}.
In this work, we also study another related observable: the boson transverse momentum dependence of charge asymmetry.
The theoretical predictions for p+p, p+Pb and Pb+Pb collisions at $5.02$~TeV are plotted in Fig.~\ref{fig:W-caspt@ppb&pbpb&pp}.
The boson rapidity region is chosen to be $|y_{c.m.}^W|<0.5$.
To have a better understanding of the distinct trends of three curves in Fig.~\ref{fig:W-caspt@ppb&pbpb&pp} we
can again resort to an analysis of the isospin effect.
In pA and AA collisions, the parton distribution asymmetry ratios could be written as
\begin{eqnarray}
\label{rud1}
r_{ud}^{pA}(x)&=&\frac{u^p(x)+u^A(x)-d^p(x)-d^A(x)}
{u^p(x)+u^A(x)+d^p(x)+d^A(x)},\nonumber \\
r_{ud}^{AA}(x)&=&\frac{u^A(x)-d^A(x)}{u^A(x)-d^A(x)}.
\end{eqnarray}
If focusing on only the isospin effect and neglecting (smaller) contributions from other CNM effects,  as discussed in the Appendix~\ref{section:appendix}, we derive some simple relations such as
\begin{eqnarray}
r_{ud}^{pA}(x)&\approx&\frac{Z}{A}r_{ud}^{pp}(x)\,\,\, ,
r_{ud}^{AA}(x)\approx \frac{Z-N}{A}r_{ud}^{pp}(x),\nonumber \\
r_{ud}^{pA}(x)&\approx&\frac{1}{2}[r_{ud}^{pp}(x)+r_{ud}^{AA}(x)],
\label{eq:r_AA}
\end{eqnarray}
and $r_{ud}^{pp}(x)\equiv r_{ud}(x)$. We can also obtain similar relations for
the ratio $r_{\bar{u}\bar{d}}(x)$.  In Fig.~\ref{fig:W-casptiso@ppb&pbpb&pp}
we plot $r_{ud}(p_{T}^W)$ and $r_{\bar{u}\bar{d}}(p_{T}^W)$ given by Eq.~(\ref{eq:r_AA}) for p+p, p+Pb and Pb+Pb, where $p_{T}^W $ is related to
the momentum fraction $x$ according to Eq.~(\ref{xpt}). Fig.~\ref{fig:W-casptiso@ppb&pbpb&pp} shows
at small $p_{T}^W$ (corresponding to small $x$) region, three curves are very close to each other.
From Fig.~\ref{fig:W-caspt@ppb&pbpb&pp} and Fig.~\ref{fig:W-casptiso@ppb&pbpb&pp}
one can see that the charge asymmetry as a function of $p_{T}^W$ might shed light on the parton distribution asymmetry ratios
$r_{ud}(x)$ and $r_{\bar{u}\bar{d}}(x)$.

For the completeness in TABLE.~\ref{table:W-yield&Raa@pbpb} we list the theoretical results of the normalized total cross section
$\langle N_{coll}\rangle^{-1}\sigma/\Delta\eta$
in the CMS final state phase space with $\Delta \eta =4.2$
at the NNLO, and the related nuclear modification factor $R_{AA}$.
CMS data are also provided for comparison. We can see that the NNLO calculations agree with the experimental data
within the experiment error bars.

\section{summary and conclusions}
\label{section:summary&conclusion}
The massive gauge boson production and decaying to a lepton pair
provide a clean process to test pQCD working at the LHC energies for both hadronic
and nuclear levels. In heavy-ion
reactions, since the final state interaction on the gauge boson production could be neglected,
the weak vector boson could be a good probe of CNM effects, and the precise comparison between the theory and
the experimental data may impose important constraints on the nuclear PDFs.

In this paper we have investigated heavy gauge bosons production
in heavy-ion collisions at the LHC at NLO and NNLO with DYNNLO program within
the framework of perturbative QCD. To consider the CNM effects, two recently published
parametrization sets of nuclear parton distribution, EPS09 and DSSZ have been used in our
simulations. We have provided the numerical results for
 the transverse momentum spectra and rapidity dependence of $Z^0$ and $W^+/W^-$,
the nuclear modification factors for $W$ and $Z^0$ particles as well as the charge asymmetry for $W$ boson productions.

For $Z^0$ production it is found that calculations at $\mathcal {O}(\alpha_s^2)$ can give a satisfactory description of the transverse momentum spectrum of $Z^0$ boson in Pb+Pb collisions recently measured by CMS Collaboration. And the theoretical predictions for $Z^0$ boson $p_T$ distribution
in p+Pb and the yield of $Z^0$ boson in p+Pb and Pb+Pb as a function of rapidity have been made.  Differences between the results with
EPS09 and DSSZ could be observed in the transverse momentum and rapidity dependence of $Z^0$ productions. It has been shown that the partonic
subprocess with at least one initial-state gluon gives dominant contribution to the $p_T$ distribution of $Z^0$ boson, and thus $Z^0$ nuclear modification factor $R(p_T^Z)$ as a function of transverse momentum is an optimal tool to study the alterations of gluon distribution in nuclear. However, the CNM effects on $Z^0$ yield as functions of the $Z^0$ boson rapidity
are dominated by the nuclear modifications on the (valence and sea) quark distributions, therefore $R(y^Z)$ may provide important information on the nuclear modifications of quark distribution functions.

The CNM effects on total $W=(W^++W^-)$ production are similar to
those in $Z^0$ production.
However, for the separate production of $W^+$ or $W^-$, the isospin effect gives the most important contribution to
the modification of $W^+$ or $W^-$ yield in both p+A collisions at $\sqrt{s_{NN}}=5.02$~TeV and Pb+Pb  at $\sqrt{s_{NN}}=2.76$~TeV
relative to that in elementary p+p reactions.
We find that the parton distribution asymmetry ratios $r_{ud}(x)$ and $r_{\bar{u}\bar{d}}(x)$ for proton could
reflect the magnitude of the isospin effect and
provide an understanding of the overall trends of  $R_{AA}(p_T^W)$ and $R_{pA}(p_T^W)$ for $W$ production.
In particular we have calculated the charge asymmetry $(N_{W^+}-N_{W^-})/(N_{W^+}+N_{W^-})$ as a function of the charged lepton
 pseudorapidity in p+Pb and Pb+Pb  collisions and compared them with the latest CMS measurement, and a good agreement between
 theory and experiment has been seen. It is interesting to show that at the same colliding energy the charge asymmetry in
 p+Pb collision may approach that in p+p collision at forward direction of the charge lepton pseudorapidity
 , but approach that in Pb+Pb at backward direction. Theoretical predictions for the charge asymmetry $(N_{W^+}-N_{W^-})/(N_{W^+}+N_{W^-})$ as a function of the transverse momentum in future heavy-ion experiments have been provided and it is shown that the charge asymmetry in p+p, p+Pb and Pb+Pb may be more pronounced in large $p_T^W$ region.

In the current study we have not included the possible effect of parton energy loss in cold nuclear matter on massive gauge boson production,
which has been found to be rather small~\cite{Neufeld:2010fj}. Moreover, here we focused on $Z^0$ and $W^{\pm}$ productions in minimum bias (MB)
nuclear collisions and have not considered the centrality dependence of gauge boson productions in heavy-ion collisions. We may delegate these discussions to other works in the future.

\begin{acknowledgments}
P.R. would like to thank Y. He, W. Dai and S.-Y. Chen for some useful discussions.
This research is supported in part by  the MOST in China under Project Nos. 2014CB845404, 2014DFG02050, and
by Natural Science Foundation of China with Project Nos. 11322546, 11221504, 11435004, and 11205024.

\end{acknowledgments}

\vspace{0.4cm}

\begin{appendix}
\section{ The isospin effect}
\label{section:appendix}
For proton, one has
\begin{eqnarray}
r_{ud}(x)&=&\frac{u(x)-d(x)}{u(x)+d(x)},\nonumber \\
\frac{u(x)}{d(x)}&=&\frac{1+r_{ud}(x)}{1-r_{ud}(x)}
=\frac{2}{1-r_{ud}(x)}-1,\nonumber \\
\frac{d(x)}{u(x)}&=&\frac{1-r_{ud}(x)}{1+r_{ud}(x)}
=\frac{2}{1+r_{ud}(x)}-1.
\end{eqnarray}
For parton distributions in nucleus, if we only consider the isospin effect and neglect contributions of other CNM effects, then
\begin{eqnarray}
u^A(x)&\approx&\frac{Z}{A}u(x)+\frac{N}{A}d(x),\nonumber \\
d^A(x)&\approx&\frac{Z}{A}d(x)+\frac{N}{A}u(x).
\end{eqnarray}
Furthermore we obtain
\begin{eqnarray}
\frac{u^A(x)}{u^p(x)}&\approx&\frac{\frac{Z}{A}u(x)+\frac{N}{A}d(x)}{u(x)}\nonumber \\
&=&\frac{Z-N}{A}+\frac{2N}{A}\frac{1}{1+r_{ud}(x)},
\end{eqnarray}
and
\begin{eqnarray}
\frac{d^A(x)}{d^p(x)}&\approx&\frac{\frac{Z}{A}d(x)+\frac{N}{A}u(x)}{d(x)}\nonumber \\
&=&\frac{Z-N}{A}+\frac{2N}{A}\frac{1}{1-r_{ud}(x)}.
\end{eqnarray}
For the parton distribution asymmetries in p+A and A+A collisions, one has
\begin{eqnarray}
r_{ud}^{pA}(x)&=& \frac{u^p(x)+u^A(x)-d^p(x)-d^A(x)}
{u^p(x)+u^A(x)+d^p(x)+d^A(x)}\nonumber \\
&\approx &\frac{Z}{A}r_{ud}(x)\equiv \frac{Z}{A}r^{pp}_{ud}(x)\nonumber  \\
\end{eqnarray}
and
\begin{eqnarray}
r_{ud}^{AA}(x)&=& \frac{u^A(x)-d^A(x)}{u^A(x)+d^A(x)}\nonumber \\
&\approx &\frac{Z-N}{A}r_{ud}(x)\equiv \frac{Z-N}{A}r^{pp}_{ud}(x).\nonumber  \\
\end{eqnarray}
And one can easily obtain the following relation as
\begin{eqnarray}
\frac{1}{2}[r_{ud}^{pp}(x)+r_{ud}^{AA}(x)]
\approx\frac{Z}{A}r^{pp}_{ud}(x)\approx r_{ud}^{pA}(x)
\end{eqnarray}
\end{appendix}

\end{document}